\documentclass[a4paper,fleqn,usenatbib]{mnras}

\usepackage{newtxtext,newtxmath}
\usepackage[T1]{fontenc}
\usepackage{ae,aecompl}

\usepackage{graphicx}	% Including figure files
\usepackage{amsmath}	% Advanced maths commands
\usepackage{amssymb}	% Extra maths symbols
\usepackage{subfig}
\title[MHD modelling of star-planet interaction]{Magnetohydrodynamic modelling of star-planet interaction and associated auroral radio emission} 

\author[S. Turnpenney et al.]{
Sam Turnpenney,$^{1}$%\thanks{E-mail: st349@le.ac.uk (KTS)}
J. D. Nichols,$^{1}$\thanks{E-mail: jdn4@leicester.ac.uk (KTS)}
G. A. Wynn$^{1}$
and X. Jia$^{2}$
\\
$^{1}$School of Physics and Astronomy, University of Leicester, Leicester, LE1 7RH, UK\\
$^{2}$Department of Climate and Space Sciences and Engineering, University of Michigan, Ann Arbor, MI, USA
}

\date{Accepted XXX. Received YYY; in original form ZZZ}

\pubyear{2020}

\begin{document}
\label{firstpage}
\pagerange{\pageref{firstpage}--\pageref{lastpage}}
\maketitle


\begin{abstract}
We present calculations of auroral radio powers of magnetised hot Jupiters orbiting Sun-like stars, computed using global magnetohydrodynamic (MHD) modelling of the magnetospheric and ionospheric convection arising from the interaction between the magnetosphere and the stellar wind. Exoplanetary auroral radio powers are traditionally estimated using empirical or analytically-derived relations, such as the Radiometric Bode's Law (RBL), which relates radio power to the magnetic or kinetic energy dissipated in the stellar wind-planet interaction. Such methods risk an oversimplification of the magnetospheric electrodynamics giving rise to radio emission. As the next step toward a self-consistent picture, we model the stellar wind-magnetosphere-ionosphere coupling currents using a 3D MHD model. We compute electron-cyclotron maser instability-driven emission from the calculated ionospheric field-aligned current density. We show that the auroral radio power is highly sensitive to interplanetary magnetic field (IMF) strength, and that the emission is saturated for plausible hot Jupiter Pedersen conductances, indicating that radio power may be largely independent of ionospheric conductance. We estimate peak radio powers of $10^{14} \; \mathrm{W}$ from a planet exposed to an IMF strength of $10^3 \; \mathrm{nT}$, implying flux densities at a distance of 15 pc from Earth potentially detectable with current and future radio telescopes. We also find a relation between radio power and planetary orbital distance that is broadly consistent with results from previous analytic models of magnetosphere-ionosphere coupling at hot Jupiters, and indicates that the RBL likely overestimates the radio powers by up to two orders of magnitude in the hot Jupiter regime. \\ \textcolor{white}{bit of blank here}
\end{abstract}


%%%%%%%%%%%%%%%%% BODY OF PAPER %%%%%%%%%%%%%%%%%%

%%%%%%%%%%%
\section{Introduction}
%%%%%%%%%%%

The direct detection of exoplanets in large parts of the electromagnetic spectrum is hindered by the high luminosity contrast ratio between the star and planet. Evidence from the Solar system planets, however, indicates that the radio waveband presents a luminosity ratio much more conducive to direct detection, with non-thermal emission from Jupiter of similar intensity to Solar radio bursts \citep{zarka2007}.  Historically, the search for exoplanetary radio emission has focussed primarily on Jupiter-like exoplanets in close orbit (3 -- 10 stellar radii) around their parent star. Such planets, given the moniker `hot Jupiters', have been targeted by several studies examining the possibility of detectable auroral radio emission \citep[e.g.][]{farrell1999, zarka2001, zarka2007, griessmeier2004, griessmeier2005, griessmeier2007, lazio2004, jardine2008, nichols2011, nichols2012candidates, hallinan2012, vidotto2017}. Although the search for radio emission has not been limited exclusively to hot Jupiters, the plausibly strong magnetic fields and intense stellar wind conditions present at these objects is thought to be favourable to the generation of radio emission through star-planet interaction. Auroral radio emission from exoplanets is envisaged to be generated by the electron-cyclotron maser instability (ECMI), the same mechanism responsible for driving radio emission from auroras at magnetized Solar system planets \citep{wu1979, treumann2006, imai2008, lamy2011}. This intense, coherent electromagnetic radiation is emitted at the local cyclotron frequency, and therefore Jupiter-like exoplanets are the prime candidates for directly detectable emission, since their potentially high intrinsic magnetic field strengths ($\sim B_{\mathrm{Jup}}$) are required to produce emission above the Earth's ionospheric cutoff frequency of $\sim$ 10 MHz.

Many previous studies estimating expected exoplanetary radio emission employ the Radiometric Bode's Law (RBL), an empirical scaling relation extrapolated from Solar system measurements between incident Poynting or kinetic energy flux and emitted radio power \citep{farrell1999, zarka2001, zarka2007, lazio2004}.  Estimates of the radio power from hot Jupiter auroras based on the RBL range between $10^{14}$ -- $10^{16}$ W, up to five orders of magnitude stronger than Jupiter's equivalent ECMI-driven emission \citep{zarka2007}, implying radio fluxes which may be detectable with the next generation of radio telescopes.  Two primary processes are assumed to mediate radio emission driven by star-planet interaction: either Alfv\'{e}n waves, such as mediates emission in the sub-Aflv\'{e}nic Io-Jupiter interaction; or magnetic reconnection, such as occurs at Earth's magnetosphere.  In considering radio emission propagated by Alfv\'{e}n waves, \citet{saur2013} and \citet{turnpenney2018} estimated total radiated energy fluxes from exoplanets of up to $10^{19} \; \mathrm{W}$, translating to radio powers of $10^{17} \; \mathrm{W}$ assuming an ECMI efficiency of 1 per cent. \citet{nichols2016}, using an analytic model, considered an Earth-type Dungey cycle process of magnetic reconnection, computing ionospheric field-aligned currents (FACs) and resulting radio powers arising from ionospheric convection.  They found that saturation of the convection-induced electric potential limited the dissipated power, and predicted auroral radio emission from hot Jupiters $\sim$2 orders of magnitude smaller than RBL-based predictions. In this paper we use a global 3D magnetohydrodynamic (MHD) model to calculate the magnetosphere-ionosphere coupling currents arising from the interaction between Sun-like stars and hot Jupiters, and hence estimate the resulting auroral radio emission.  By using a numerical global MHD model, this study extends the analytic work of \citet{nichols2016} to enable a self-consistent calculation of the FACs at exoplanets. The numerical model, as used in this study, takes a set of input boundary conditions and generates a 2D map of the ionospheric FAC density distribution, from which radio power is calculated in post-processing.   The overall power is expected to be strongly influenced externally, by the interplanetary magnetic field (IMF) strength, and internally, by the ionospheric Pedersen conductance.  Therefore, we run simulations across a broad range of these two parameters, comparing the results with those of \citet{nichols2016}.

This paper begins with an overview of the theoretical background relevant to the MHD model, along with the formulation used to compute radio powers from the field-aligned current output of the model. Then follows a description of the method employed, before results of the modelling work are presented and discussed. A case intermediate between Earth and hot Jupiter exoplanets is first studied, before cases more appropriate to exoplanets are examined.  

%%%%%%%%%%%%%%%%
\section{Theoretical background}
%%%%%%%%%%%%%%%%

%%%%%%%%%%%%%%%%%%%%%
\subsection{Magnetohydrodynamic model}
%%%%%%%%%%%%%%%%%%%%%

Global 3D MHD simulations, based on first principle physics,  are a powerful tool for modelling the dynamics and evolution of magnetic fields and plasma flows in space weather and astrophysical phenomena.  In use since the 1980s, early models relied on techniques such as finite-difference methods to solve a system of discretized MHD equations \citep{vanleer1979, leboeuf1981, wu1981, brecht1982}. Computational solutions of MHD equations require discretization of the MHD equations, which inherently introduces errors into the solution. Modern MHD models use advanced solution techniques which improve the efficiency of the solution and minimise such discretization errors \citep{gombosi2004}. Such methods rely on approximations to solve a Riemann problem, the form of initial value problem presented in MHD numerical analysis over a finite volume.  The 3D MHD solver used in this work is the `Block Adaptive Tree Solarwind Roe Upwind Scheme' (BATS-R-US) software first outlined by \citet{powell1999}, and developed at the University of Michigan. The computational scheme of BATS-R-US is based on the same elements used in many state-of-the-art MHD codes, and this section describes each of those elements in the scheme.

A governing set of 3D ideal MHD equations is first defined. Various forms of these equations are expressible, and the form chosen is dictated by factors which will ultimately aid in the computational solution of these equations.  The set of ideal MHD equations used in BATS-R-US are expressed in a gasdynamic conservative form \citep{gombosi2004} 

\begin{equation} \label{massconv}
\frac{\partial \rho}{\partial t} + \nabla \cdot (\rho \boldsymbol{u}) = 0
\end{equation}
\begin{equation} \label{momconv}	
 \frac{\partial (\rho \boldsymbol{u})}{\partial t}  + \nabla \cdot (\rho \boldsymbol{u u} + p \textbf{I})= \frac{1}{\mu_0}  \boldsymbol{j} \times \boldsymbol{B}
\end{equation}
\begin{equation} \label{fara}
\frac{\partial \boldsymbol{B}}{\partial t} + \nabla \times \boldsymbol{E} = 0
\end{equation}
\begin{equation} \label{energyeq}	
\frac{\partial E_{\mathrm{gd}}}{\partial t} + \nabla \cdot \left[ \boldsymbol{u} (E_{\mathrm{gd}} + p) \right] = \frac{1}{\mu_0} \boldsymbol{u} \cdot ( \boldsymbol{j} \times \boldsymbol{B}),
\end{equation}
where $\textbf{I}$ is the identity matrix and the total gasdynamic energy $E_{\mathrm{gd}}$, is given by

\begin{equation}\label{energyeq2}
E_{\mathrm{gd}} = \frac{1}{2} \rho u^2 + \frac{1}{\gamma - 1} p
\end{equation}
where $\gamma$ is the ratio of the specific heats.  This equation set contains an expression for the conservation of mass (equation \ref{massconv}), conservation of momentum (equation \ref{momconv}), an expression of Faraday's law (equation \ref{fara}), and an energy equation (\ref{energyeq}). These partial differential equations are manipulated into a non-dimensional, symmetrizable form, the full details of which can be found in \citet{powell1999}. A computational domain is divided into cells over which the MHD equations are integrated. From the symmetrizable form \citet{powell1994} showed that it is possible to derive a Roe-type approximate Riemann solver for the 3D equations.  First described by \citet{roe1981}, this is a method for solving partial differential equations by estimating the flux at the interface between two computation cells in some finite-volume discretized domain.  Such solvers are required in magnetohydrodynamics, since iterative solutions are costly, and therefore approximations must be made. 

The computation domain is divided into a grid of Cartesian cells, and the cells are structured into blocks typically consisting of between $4\times4\times4$ and $12\times12\times12$ individual cells.  The block-adaptive technology of BATS-R-US allows the computational grid to be adapted based on prespecified physical criteria, such that blocks can be refined in regions where interesting physical features emerge.  Adaptive mesh refinement is extremely effective when the problem being treated contains disparate length scales, and also removes any initial grid-based bias in the solution.

The Space Weather Modelling Framework (SWMF) is a software package which integrates several different physics domains extending from the solar surface to the planetary upper altmosphere \citep{toth2005}. BATS-R-US is used in a number of these components where MHD solutions are required, and physically meaningful combinations of components can be coupled together to study a wide variety of space weather events and phenomena \citep{toth2012}. While developed originally to study Sun-Earth events, the SWMF has since been adapted and applied to other solar system planets, satellites and comets \citep[e.g.][]{jia2019, jia2016, toth2016, huang2016}, and may reasonably be used for the study of extrasolar astrophysical systems where the physics domains are appropriate, with some adaptation where may be required.

%%%%%%%%%%%%%%%%%%%%%%%%%
\subsection{Field-aligned current and radio power}
%%%%%%%%%%%%%%%%%%%%%%%%%

This study utilizes the Global Magnetosphere (GM) component of the SWMF, coupled with the Ionospheric Electrodynamics (IE) component \citep{ridley2004}. The GM domain constructs the magnetic environment and plasma dynamics around the planet, and contains features such as the bow shock, magnetopause, and magnetotail.  Upstream boundary condition for the GM component can be obtained from coupling with the Inner Heliosphere component of the SWMF, but in this work are simply input into the model based on reasonable values, as will be discussed below. Currents from the GM component are mapped down along magnetic field lines to provide the field-aligned current boundary conditions for the IE component.  The domain of the IE component is a height-integrated spherical surface. Formally, this component is  a two-dimensional electric potential solver, which computes conductances and particle precipitation from FACs. The process can be summarized as follows: 1) Field-aligned currents are calculated by $\nabla \times \boldsymbol{B}$ at 3.5 $R_{\mathrm{P}}$, a value also employed by \citet{ridley2004}, where $\boldsymbol{B}$ is the local magnetic field; 2) The currents are then mapped down along field lines to a nominal ionospheric altitude of $\sim 110 \; \mathrm{km}$ using the planetary dipolar field, and are scaled by the ratio $B_{\mathrm{I}} / B_{3.5}$, where $B_{\mathrm{I}}$ and $B_{3.5}$ are the magnetic field strengths at the ionosphere and 3.5 planetary radii respectively. 3) Next, a height-integrated ionospheric conductance map is generated and the electric potential is calculated, which is then mapped out along magnetic field lines to the simulation's inner boundary at 2.5 $R_{\mathrm{P}}$, where flow velocities and electric fields are prescribed \citep{ridley2001, ridley2004}.

Of the several variables output from the IE component, this work is principally concerned with the ionospheric FAC density $j_{||}$, and the cross-polar cap potential (CPCP) $\Phi_{\mathrm{CPC}}$ outputs. By integrating the total upward or downward FAC density output from the IE component over one hemisphere, the total current, $I_{\mathrm{tot}}$, flowing into or out of the ionosphere is obtained by

\begin{equation} \label{itot}
I_{\mathrm{tot}} = \int^{2 \pi}_0 \int^{\pi/2}_0 R_{\mathrm{P}}^2 j_{||} \sin \theta \mathrm{d}\theta \mathrm{d}\phi, 
\end{equation}
where $\theta$ and $\phi$ are the conventional spherical coordinates of colatitude and azimuth respectively. Total auroral power is also calculated in post-processing by integrating precipitating electron energy flux $E_{\mathrm{f}}$ over one hemisphere:

\begin{equation} \label{pow-swmf}
P_{\mathrm{tot}} = \int^{2 \pi}_0 \int^{\pi/2}_0 R_{\mathrm{P}}^2 E_{\mathrm{f}}(\theta, \phi) \sin \theta \mathrm{d} \theta \mathrm{d}\phi.
\end{equation}
Qualitatively, the precipitating electron energy flux is the kinetic energy carried by the downward-flowing electrons associated with the upward field-aligned current.  The maximum FAC which can be carried by unaccelerated electrons in an isotropic Maxwellian velocity space distribution is given by

\begin{equation} \label{j0}
j_{\|i0} = e n \left(\frac{W_{\mathrm{th}}}{2 \pi m_{\mathrm{e}}} \right)^{1/2},
\end{equation} 
where $e$ and $m_{\mathrm{e}}$ are the electron charge and mass respectively, and $W_{\mathrm{th}}$ and $n$ are are the thermal energy and number density of the magnetospheric electron source population respectively, for which we employ canonical jovian values established from Voyager measurements of $W_{\mathrm{th}} = 2.5 \; \mathrm{keV}$ and $n = 0.01\; \mathrm{cm}^{-3}$ throughout this work \citep[e.g.][]{scudder1981}. We note that these values may differ significantly at hot Jupiters, although no reliable estimates exist at present. The source plasma number density is in relation to the evacuated auroral field lines, and is therefore expected to be much reduced from the ambient plasma density. Qualitatively, the effect of varying these parameters, however, will be to increase the precipitating electron energy flux and auroral power with increasing plasma thermal energy, and with decreasing plasma density.  The jovian values are employed here in the absence of any information for realistic values at hot Jupiters, although future work may investigate a range of these parameter values to determine quantitatively the effect on auroral radio emission. In general, the FACs will be larger than can be carried solely by unaccelerated electrons, and therefore must be driven by a field-aligned electric potential.  In common with previous works computing intense exoplanetary and ultracool dwarf auroral radio emissions \citep{nichols2011, nichols2012candidates, nichols2012, turnpenney2017} , we employ \citeauthor{cowley2006}'s (\citeyear{cowley2006}) relativistic extension of \citeauthor{knight1973}'s (\citeyear{knight1973}) current-voltage relation for parallel electric fields given by

\begin{equation}
\left( \frac{j_{\|i}}{j_{\|i0}} \right) = 1 + \left(\frac{e \Phi_{\mathrm{{min}}}}{W_{\mathrm{th}}}  \right) + \frac{\left(\frac{(e \Phi_{\mathrm{min}}}{W_{\mathrm{th}}}\right)^2}{2 \left[ \left(\frac{m_{\mathrm{e}} c^2}{W_{\mathrm{th}}} \right)+ 1\right]},
\end{equation}
where $\Phi_{\mathrm{min}}$ is the minimum field-aligned voltage required to drive the current $j_{\|i}$ in the ionosphere, and $c$ is the speed of light in a vacuum. The corresponding precipitating electron energy flux is given by 

\begin{equation}
\frac{E_{\mathrm{f}}}{E_{\mathrm{f}0}} = 1 +\left( \frac{e\Phi_{\mathrm{min}}}{W_{\mathrm{th}}} \right) + \frac{1}{2}  \left( \frac{e\Phi_{\mathrm{min}}}{W_{\mathrm{th}}} \right)^2 + \frac{ \left( \frac{e\Phi_{\mathrm{min}}}{W_{\mathrm{th}}}  \right)^3 }{ 2 \left[ 2 \left( \frac{m_{\mathrm{e}} c^2}{W_{\mathrm{th}}} \right) + 3 \right]},
\end{equation}
where $E_0$ is the maximum unaccelerated electron energy flux, corresponding to equation (\ref{j0}), given by 

\begin{equation}
E_{\mathrm{f}0} = 2 n W_{\mathrm{th}} \left( \frac{W_{\mathrm{th}}}{2 \pi m_{\mathrm{e}}} \right)^{1/2}.
\end{equation}
Assuming, in common with observations of Jupiter and Saturn, an ECMI efficiency of 1 per cent \citep{gustin2004, clarke2009, lamy2011}, the emitted auroral radio power is

\begin{equation} \label{rad-pow}
P_\mathrm{r} = \frac{P_{\mathrm{tot}}}{100}.
\end{equation}
Finally, the spectral flux density is calculated by

\begin{equation} \label{flux}
F_{\mathrm{r}} = \frac{P_{\mathrm{r}}}{1.6 s^2 \Delta \nu},
\end{equation}
where $s$ is the distance to the exoplanetary system from the Earth, $\Delta \nu$ is the radio emission bandwidth, and a beaming angle of 1.6 sr is assumed on the basis of Jupiter's observed ECMI emission beaming angle \citep{zarka2004}. Since cyclotron maser emission is generated at the local electron-cyclotron frequency, the bandwidth is the difference between the magnetic field strength at the location of the field-aligned potential and the ionosphere. This formulation assumes that the potential is located high enough up the field line from the ionosphere that the field strength there is much smaller than the field strength in the ionosphere. This assumption is valid beyond a few planetary radii owing to the $r^{-3}$ dependence for a dipole planetary field. We therefore assume that the bandwidth $\Delta \nu$ is equal to the electron cyclotron frequency in the polar ionosphere and hence given by

\begin{equation}
\Delta \nu = \frac{e B_i}{2 \pi m_{\mathrm{e}}},
\end{equation}
where $B_i$ is the ionospheric magnetic field strength. \\

%%%%%%%%%%%
\section{Methodology}
%%%%%%%%%%%

Input parameters were chosen for the SWMF to simulate a magnetised hot Jupiter interacting with the IMF and stellar wind of a solar-type star.  A planet of Jupiter mass ($1.9 \times 10^{27} \, \mathrm{kg}$), radius ($71,492 \, \mathrm{km}$), and equatorial magnetic field strength ($426,400\,\mathrm{nT}$) was specified, with a dipole magnetic field aligned with the planetary spin axis. The orientation of the magnetic field is opposite to that of Jupiter, i.e. the planetary field is pointing northward at the equator.  Each simulation was run with a planetary plasma density of $10^7 \, \mathrm{cm^{-3}}$ , and a temperature of $8,000\, \mathrm{K}$ at the inner boundary of the simulation, values consistent with those based on modelling of atomic hydrogen, heavy atoms and ions surrounding hot Jupiters \citep{munoz2007, koskinen2013, shaikhislamov2016}. The incident plasma velocity of the simulations was set to $250\, \mathrm{km\,s^{-1}}$. This value represents the impinging plasma flow velocity taking into account both the stellar wind outflow and the Keplerian orbital velocity of the planet in the hot Jupiter regime.  We note that although in reality the incident plasma velocity will be predominantly azimuthal due to the high orbital velocity of planet, in this work this incident plasma velocity is prescribed as an input in the model, such that in the results that follow (i.e. Figs 1 and 2) the X-axis is aligned with the incident plasma velocity.  This transposition is purely for convenience of modelling, and does not affect the resulting auroral radio power calculations.

Since the stellar wind plasma temperature close to the star in the orbital distance regime of hot Jupiters is approximately the coronal value, a stellar wind plasma temperature input of $2\, \mathrm{MK}$ was employed throughout this study. The stellar wind density was set to $10^4\, \mathrm{cm^{-3}}$ in all simulations, a value appropriate for the hot Jupiter regime based on analytic modelling by \citet{nichols2016}.

In the work presented here, simulations were run for an entirely southward IMF orientation, with $B_{\mathrm{sw}}$ values ranging from 1-- $10^6\,\mathrm{nT}$, and with a constant ionospheric Pedersen conductance between 1 -- $10^4 \, \mathrm{mho}$. The assumption of a constant Pedersen conductance provides a reasonable first-order approximation to the global average. A uniform zero Hall conductance was used throughout, which along with a constant non-zero Pedersen conductance forms the simplest ionospheric conductance model to approximate a realistic magnetospheric configuration. Such a configuration has been used as a standard ionospheric model in many previous MHD simulations \citep[e.g.][]{fedder1987, jia2012driving, jia2012magnetospheric}. 

The simulations were run on a 3D Cartesian computational grid of $256 \,\times 256 \, \times 256 \, R_{\mathrm{P}}$. Although BATS-R-US may be run in a time-accurate mode, since this work focuses on hypothetical events, we instead used iterative local time stepping, in which each computation cell takes different time steps and the simulation progresses for a fixed number of iterations to converge on a steady-state solution. At the inner boundary of each simulation the grid was initially highly resolved near the planet with cells of 1/8 $R_{\mathrm{P}}$ in size, while the remainder of the grid is incrementally more coarsely resolved moving out further from the planet. This initial resolution is entirely geometric, i.e. it is not based on any physical criteria, but rather is determined based on expectations of where interesting regions of the solution will emerge requiring high resolution. Each simulation was allowed to run for 3,000 iterations before the grid was refined using the adaptive mesh refinement facility within the MHD code.  Refinement added 10\% more cells in regions of large $\nabla P$ and $\nabla \times \boldsymbol{B}$ before the simulation resumed running.  Refinement based on these criteria was performed every 300 iterations up to a total of 6,000 iterations, giving a final grid containing approximately 15 million cells.  After the final refinement the simulation was then allowed to run for a total of 50,000 iterations, by which point the solution had reached an approximate steady-state. 

To validate the approach described above, the model was first tested by replicating earlier results of \citet{ridley2004} and \citet{ridley2007}, the details of which can be found in Appendix \ref{appA}. This work builds on those studies with vastly increased IMF strengths and Pedersen conductances. 

%%%%%%%%%%%
\section{Results}
%%%%%%%%%%%

%%%%%%%%%%%%%%%%%%%%
\subsection{Magnetospheric structure}
%%%%%%%%%%%%%%%%%%%%

\begin{figure*}
    \centering
    \includegraphics[scale=0.8]{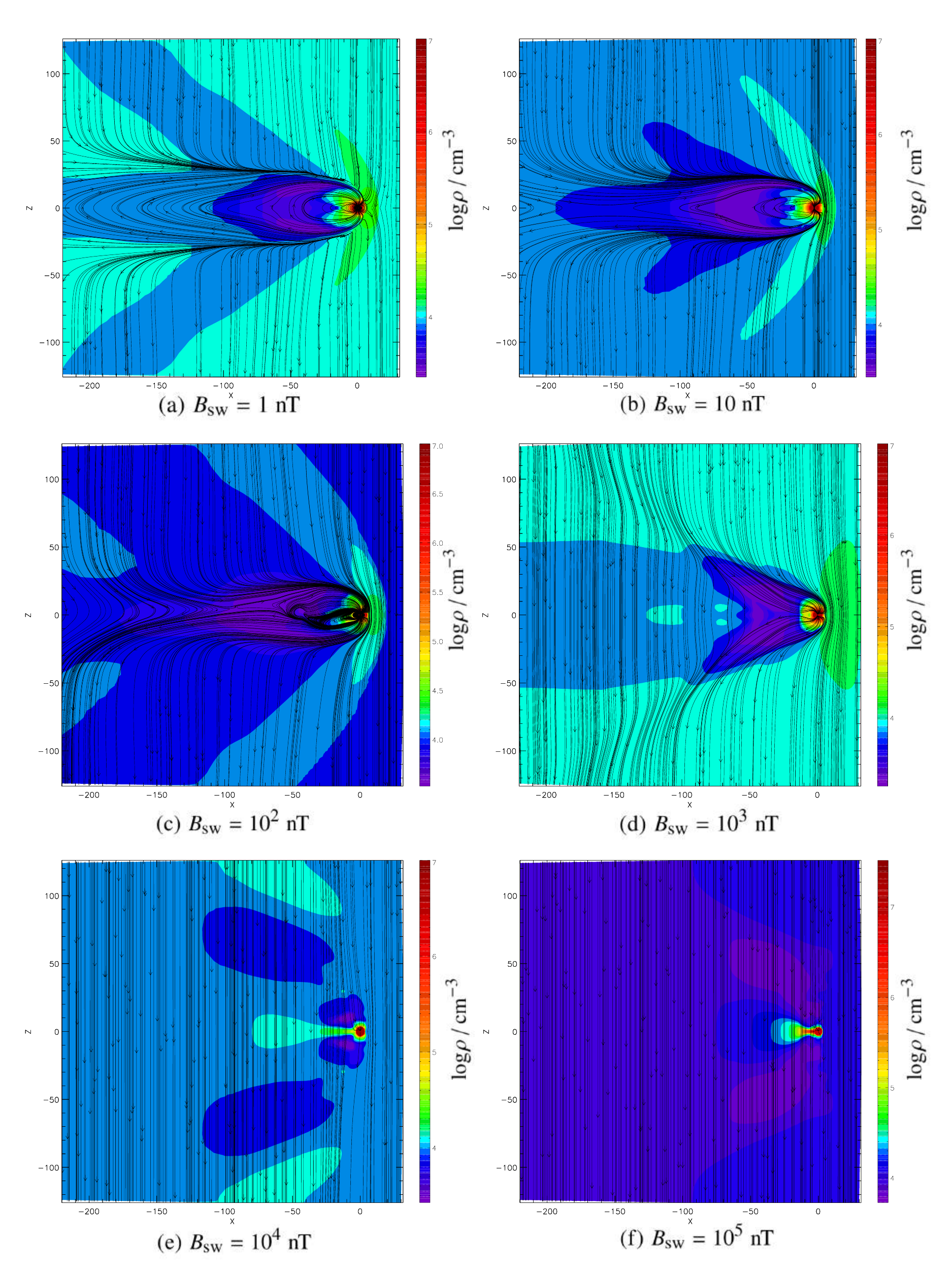}
    \caption{Plasma density with traces showing magnetic field lines in the $X$-$Z$ plane (in GSM co-ordinates).  The planet is situated at $Z = X =0$ in each plot, and the stellar wind is flowing from right to left.  The co-ordinates are given in units of planetary radii. Note that the density scale differs between the individual plots.}
    \label{fig:magfields}
\end{figure*}

\begin{figure*}
    \centering
    \includegraphics[scale=0.8]{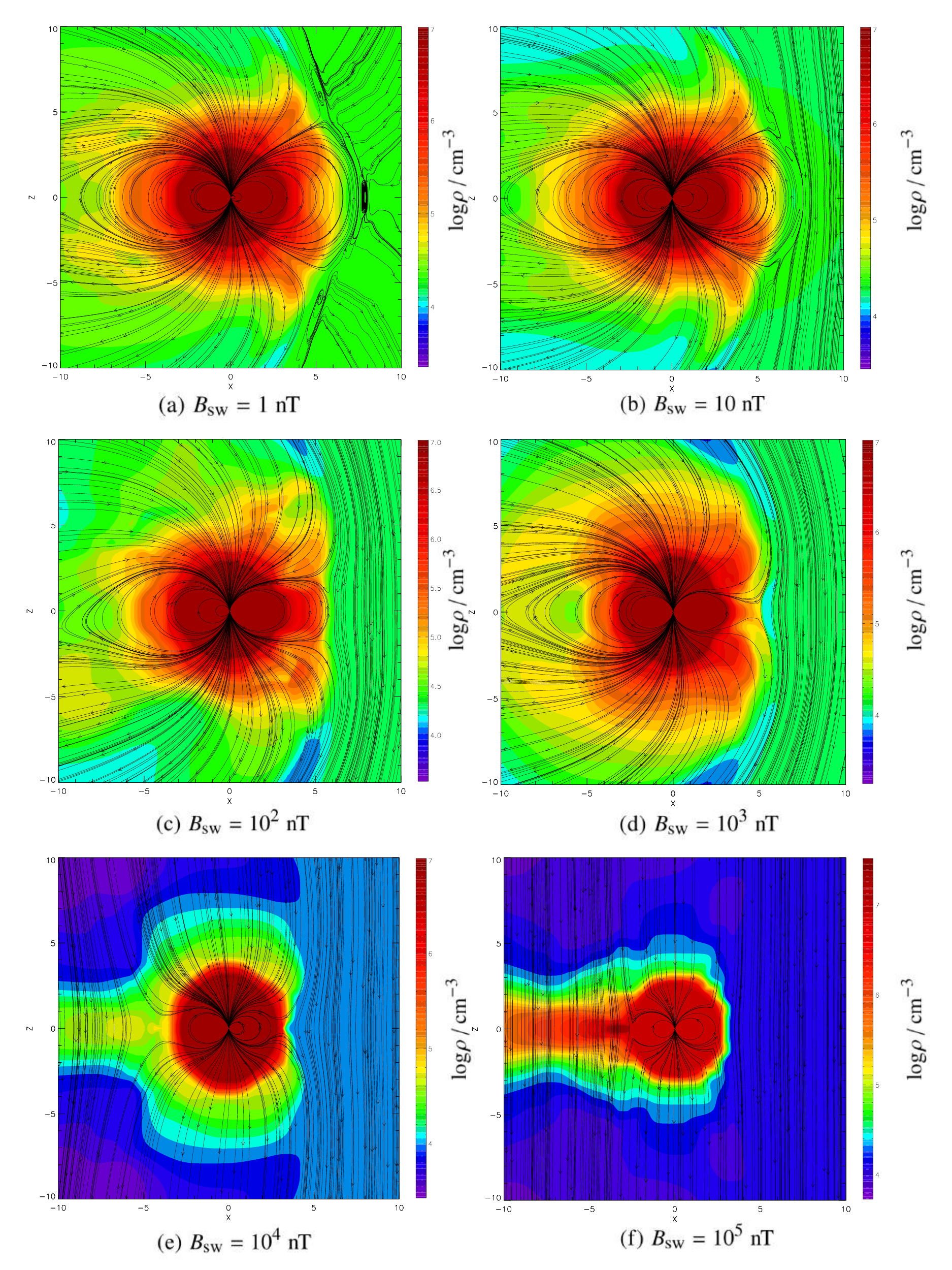}
    \caption{As for Fig. \ref{fig:magfields} but for a smaller scale to show the magnetospheric topology in close proximity to the planet}
    \label{fig:magfieldszoomed}
\end{figure*}

A series of simulation results of the magnetospheric field morphology and plasma density is shown in Fig. \ref{fig:magfields} for IMF $B_{\mathrm{sw}}$ values ranging from 1 -- $10^5\, \mathrm{nT}$.  In each run the Pedersen conductance was initially $\Sigma_{\mathrm{P}} = 10^4 \, \mathrm{mho}$, representing the highly conductive ionospheres expected at hot Jupiters. The remainder of the input parameters were as stated in the previous section.  The magnetic field lines are traced in the $Y = 0$ plane, with the incident plasma flowing from right to left. Note that in these plots the field line spacing does not necessarily represent magnetic field strength. The lower end of the IMF range represents conditions analogous to the IMF experienced at the Earth. In the region typically associated with hot Jupiters, i.e. 3 -- 10 stellar radii, an IMF of approximately $10^3$  -- $10^4 \, \mathrm{nT}$ is expected \citep{nichols2016}. Higher IMF strengths have been examined which represent either planets orbiting the star extremely closely ($<3$ stellar radii), or planets orbiting stars with exceptionally strong magnetic fields. For instance, at a distance of $10 \; R_*$, assuming a predominantly radial field such that $B_{\mathrm{sw}} \propto 1 /r^2$, a $10^6$ nT IMF equates to a star with a surface magnetic field strength of $10^8$ nT, approximately three orders of magnitude great than the solar magnetic field.

Fig. \ref{fig:magfields}(a) shows that when the planet is exposed to $B_{\mathrm{sw}} = 1 \,\mathrm{nT}$ the magnetosphere formed is similar to that at the Earth, i.e. with a clearly visible magnetotail and an apparent magnetosphere on the sub-stellar wind side of the planet. As the IMF is increased $B_{\mathrm{sw}} = 10^2\,\mathrm{nT}$ and $B_{\mathrm{sw}} = 10^3\,\mathrm{nT}$, Figs. \ref{fig:magfields}(c) and (d) show the lobes of the tail opening as the upstream Alfv\'{e}n Mach number becomes lower, and the formation of Alfv\'{e}n wings draped across the planet becomes apparent.  At $B_{\mathrm{sw}} = 10^4\, \mathrm{nT}$, the flow has become sub-Alfv\'{e}nic, and the Alfv\'{e}n wings are formed at a large angle from the equatorial plane. As the IMF is increased further to $B_{\mathrm{sw}} = 10^5 \, \mathrm{nT}$ the planetary field is dwarfed by the stellar field, and essentially presents no perturbation to the overwhelming IMF. 

Fig. \ref{fig:magfieldszoomed} shows the magnetospheric field line topology and plasma density in close proximity to the planet for the same simulations as Fig. \ref{fig:magfields}.  These plots reveal the compression of the sub-stellar wind side magnetosphere due to the IMF pressure as $B_{\mathrm{sw}}$ is increased, as well as escape of the dense planetary plasma along open field lines. Note that in Figs. \ref{fig:magfieldszoomed}(e) and (f) the substellar magnetosphere has collapsed below the $2.5 \, R_{\mathrm{p}}$ inner boundary of the simulation.

%%%%%%%%%%%%%%%%%%%%%%%%%%%%%
\subsection{Ionospheric Electrodynamics} \label{iono-elec} 
%%%%%%%%%%%%%%%%%%%%%%%%%%%%%

\begin{figure*}%[H]
     \subfloat[Ionospheric field-aligned current density in units of $\upmu \mathrm{A} \, \mathrm{m}^{-2}$\label{subfig-1:dummy}]{%
       \includegraphics[width=0.45\textwidth]{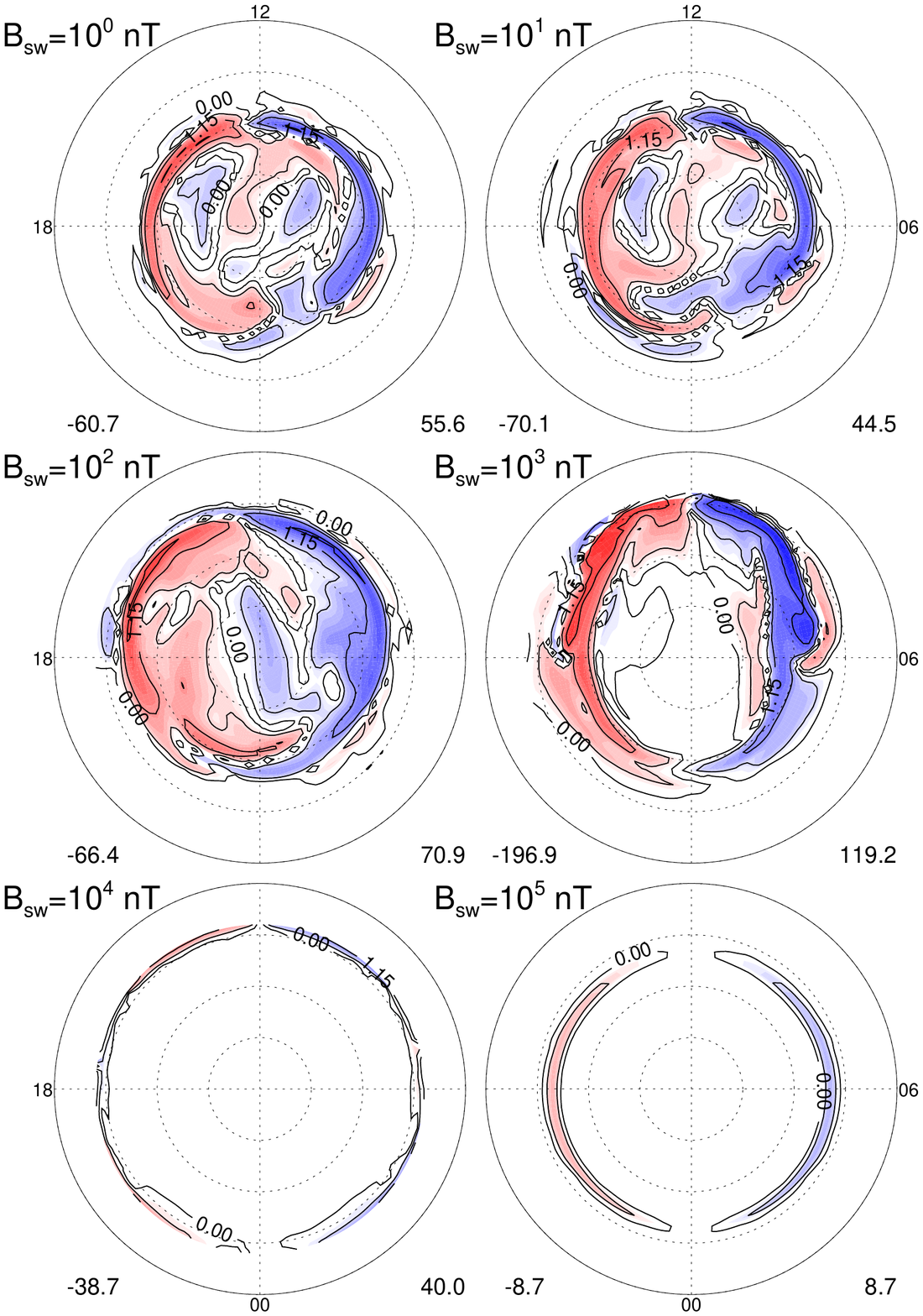}
     }
     \hfill
     \subfloat[Ionospheric cross-polar cap potential in units of kV \label{subfig-2:dummy}]{%
       \includegraphics[width=0.45\textwidth]{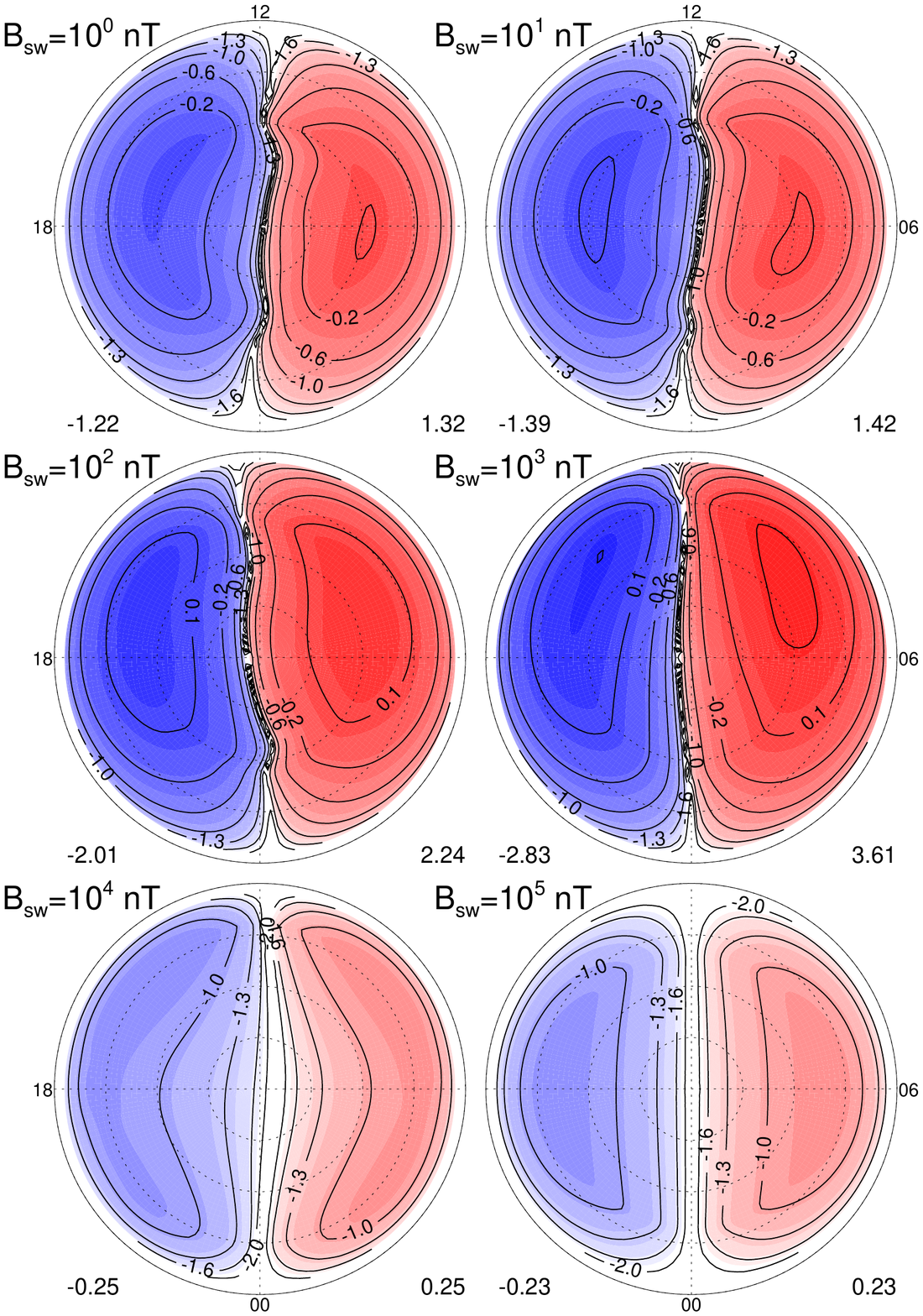}
     }
     \caption{Ionospheric maps of (a) field-aligned current density and (b) cross-polar cap potential for a fixed ionospheric Pedersen conductance of $\Sigma_{\mathrm{P}} = 10^4 \; \mathrm{mho}$, and $B_{\mathrm{sw}}$ ranging from 1 - 10$^5 \, \mathrm{nT}$. The minimum and maximum values of the two parameters are indicated below each plot. In each case, contours of absolute values are shown on a logarithmic scale, with negative values coloured blue and positive values coloured red.}
     \label{fig:iono_sigconst2}
   \end{figure*}

Fig. \ref{fig:iono_sigconst2} shows a set of ionospheric plots of FAC density and CPCP at a constant Pedersen conductance value of $\Sigma_{\mathrm{P}} = 10^4 \; \mathrm{mho}$, and with $B_{\mathrm{sw}}$ range of $1 - 10^5\; \mathrm{nT}$. In each FAC density plot positive upward current is indicated in red, and negative downward current by blue. The morphology and magnitudes of both quantities within these plots vary as $B_{\mathrm{sw}}$ is increased.  At low IMF strengths the auroral FAC density is somewhat irregular and diffuse in shape, and is situated at higher colatitudes of $\sim 20-30^{\circ}$, slightly displaced towards the sub-stellar wind side of the planet. The upward FAC density peaks for $B_{\mathrm{sw}} = 10^3 \; \mathrm{nT}$ at $199  \; \upmu \mathrm{A} \, \mathrm{m^{-2}}$. At $B_{\mathrm{sw}} = 10^4 \; \mathrm{nT}$ the auroral FAC structure is a narrow oval at $\sim 30^{\circ}$ colatitude, with the peak upward FAC density of $39.97 \; \upmu \mathrm{A} \, \mathrm{m^{-2}}$.  As $B_{\mathrm{sw}}$ is increased to $10^5$ the magnitude of the upward current density falls significantly to 8.67 $\upmu\mathrm{A} \,\mathrm{m^{-2}}$.

A similar saturation and turnover is seen in the CPCP results in Fig. \ref{fig:iono_sigconst2}(b). From a CPCP potential of $\sim 3 \; \mathrm{keV}$ for  $B_{\mathrm{sw}} = 1\; \mathrm{nT}$, the potential peaks around $6 \;\mathrm{keV}$ at  $B_{\mathrm{sw}} = 10^3 \; \mathrm{nT}$, before a substantial drop to $\Phi_{\mathrm{CPC}} = 0.457 \; \mathrm{kV}$ at $B_{\mathrm{sw}}= 10^5 \; \mathrm{nT}$.  Saturation of the CPCP, first discussed by \citet{hill1976}, was subsequently tested by \citet{siscoe2002} against results from MHD simulations, and was modelled in terms of the stellar wind parameters. Various interpretations have been offered for the phenomenon of CPCP saturation. \citet{siscoe2002} interpreted the saturation as a weakening of the planetary field at the magnetopause due to the field arising from region 1 currents, thus limiting the rate at which reconnection occurs on the sub-stellar wind side of the planet.  Alternatively, \citet{kivelson2008} argued that saturation occurs when the solar or stellar wind impedance is greater than the ionospheric impedance, i.e. when the Pedersen conductance $\Sigma_{\mathrm{P}}$ dominates the Alfv\'{e}n conductance $\Sigma_{\mathrm{A}}$, causing a partial reflection of Alfv\'{e}n waves incident on the ionosphere from the solar or stellar wind. The available magnetospheric convection potential is given by \citep{nichols2016}

\begin{equation}
\Phi_{\mathrm{m}} = \chi R_{\mathrm{mp}} E_{\mathrm{sw}}, \label{phitot}
\end{equation}
where $E_{\mathrm{sw}}$ is the stellar wind motional electric field, and $\chi$ is the fraction of the magnetopause standoff distance $R_{\mathrm{mp}}$ which constitutes the stellar wind reconnection channel. Observations for Earth determine a value of  $\chi \approx 0.5$ \citep{milan2004}, and this value is therefore also employed here. In the reflected Alfv\'{e}n wave interpretation of \citet{kivelson2008}, the electric potential across the ionosphere is given by  
\begin{equation}
\Phi_{\mathrm{CPC}} = \frac{2 \xi \Phi_{\mathrm{m}} \Sigma_{\mathrm{A}}}{\Sigma_{\mathrm{P}} + \Sigma_{\mathrm{A}}}, \label{CPCP-nichols}
\end{equation}   
where the width of the interaction channel, specified by  \citet{kivelson2008} as $0.1 \pi R_{\mathrm{mp}}$ is accounted for by the factor $\xi =(0.1\pi \big/ \chi)$.  Hence, when $\Sigma_{\mathrm{P}} \gg \Sigma_{\mathrm{A}}$ the CPCP tends towards 

\begin{equation}
\Phi_{\mathrm{sat}} = \frac{2 \xi \Phi_{\mathrm{m}} \Sigma_{\mathrm{A}}}{\Sigma_{\mathrm{P}}}
\end{equation}
and saturation occurs.  For a fixed Pedersen conductance, increased IMF strength leads to a reduced Alfv\'{e}n conductance, and thus a decrease in CPCP beyond saturation as observed in the results in Fig. \ref{fig:iono_sigconst2}(b). The saturation effect is also influenced by the decreasing magnetopause standoff distance, but the dominant contributing factor is decreased Alfv\'{e}n conductance, since the sub-stellar wind side magnetosphere is completely eroded under high IMF strengths.

% These plots are from lhsXtimestep0.4 runs (copied to timestep0.4X in order to produce them like this). So they are for Sigp = 1e4 mho. These pdfs need to be edited in order to indicate the Bsw values for each 

A notable feature of the plots in Figs. \ref{fig:iono_sigconst2}(a) is the absence of strong region 2 currents.  This is an artefact of the MHD code, with several possible causes suggested by \citet{ridley2001}. Since region 2 currents are generated close to the inner boundary of the model, a high resolution is required to produce currents that are even a fraction of the region 1 currents. Increasing the resolution would increase the time taken for simulations to run, and is therefore a trade-off that must be made in consideration of running the models in a timely fashion.  Another option which should achieve the same result is to move the inner boundary to a lower value (e.g. from $2.5 \,R_{\mathrm{P}}$ to $1.5\, R_{\mathrm{P}}$).  As with increasing the resolution, this solution also increases the run time of the simulations.  Another possible cause of the weak region 2 currents is that gradient and curvature drifts of particles at different energies is not addressed by BATS-R-US. The pressure gradient that results from the reconfiguration of the plasma by these drifts may be a source of region 2 currents.

%%%%%%%%%%%%%%%%%%%%%%%%%%%%%%%%%%%%%%%%%%%%%%
\subsection{Responses to variable IMF strength and Pedersen conductance} \label{response} 
%%%%%%%%%%%%%%%%%%%%%%%%%%%%%%%%%%%%%%%%%%%%%%

\begin{figure}
    \centering
    \includegraphics[width=1.0\linewidth]{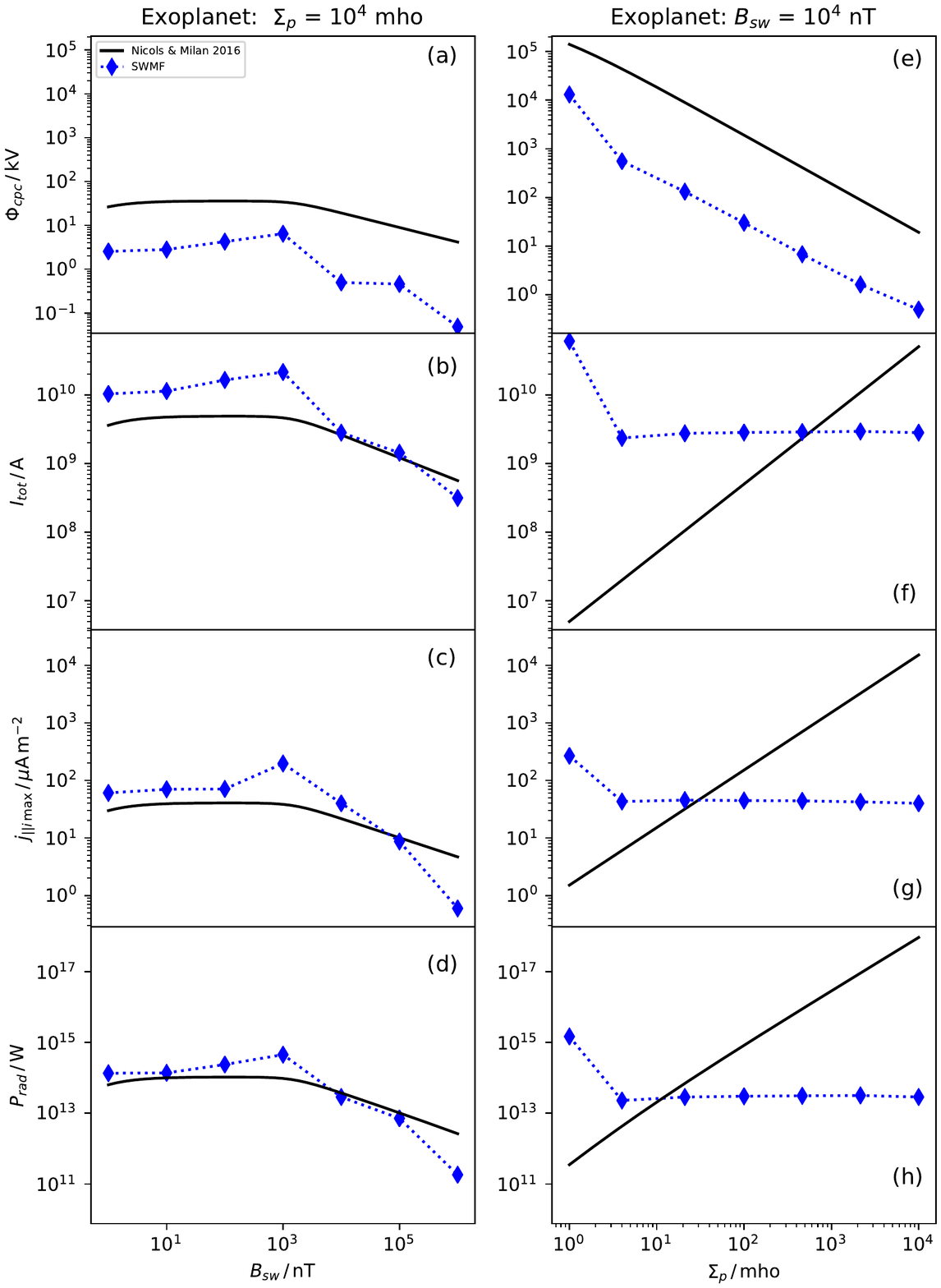}
    \caption[STILL NEEDS THE RHS PLOTS.  ALSO TRY AND GET UNCERTAINTIES WITH e.g. MAX AND MIN VALUES FROM SAY FINAL 10k ITERATIONSCross-polar cap potential, total FAC, maximum FAC density, and radio power versus IMF strength for $\Sigma_{\mathrm{P}} = 10^4 \; \mathrm{mho}$ and $B_{\mathrm{sw}} = 10^4 \; \mathrm{nT}$]{(a) Cross-polar cap potential, (b) total field-aligned current, (c) maximum ionospheric field-aligned current density, and (d) auroral radio power as functions of IMF $B_{\mathrm{sw}}$ for an exoplanet with ionospheric Pedersen conductance of $\Sigma_{\mathrm{P}} = 10^4 \; \mathrm{mho}$.  (e) Cross-polar cap potential, (f) total current, (g) maximum field-aligned current density, and (h) auroral radio power as functions of Pedersen conductance for an exoplanet exposed to $B_{\mathrm{sw}} = 10^4 \; \mathrm{nT}$.  Blue lines represent the SWMF results, with diamonds denoting the values are which MHD simulations were conducted. Black lines show the analytic results using the model of \citet{nichols2016}.}
    \label{fig:stacked_plots_exo_lhs_rhs}
\end{figure}

To fully understand the response of the global simulations to the key driving factors, Fig. \ref{fig:stacked_plots_exo_lhs_rhs} shows plots of CPCP, total current, maximum ionospheric FAC density, and radio power, versus both $B_{\mathrm{sw}}$ and Pedersen conductance. In Figs. \ref{fig:stacked_plots_exo_lhs_rhs}(a) -- (d) the Pedersen conductance was fixed at $\Sigma_{\mathrm{P}} = 10^4 \; \mathrm{mho}$, since \citet{koskinen2010} showed that  hot Jupiters likely possess highly conductive ionospheres ($\Sigma \approx 10^4$ -- $10^5 \; \mathrm{mho}$), and the modelled parameters are plotted as a function of $B_{\mathrm{sw}}$ from 1 -- $10^6\; \mathrm{nT}$. In Fig. \ref{fig:stacked_plots_exo_lhs_rhs}(e) -- (h) a fixed IMF value of $B_{\mathrm{sw}} = 10^4 \; \mathrm{nT}$ was used, and $\Sigma_{\mathrm{P}}$ was varied.  The modelled CPCP, shown in Fig. \ref{fig:stacked_plots_exo_lhs_rhs}(a), initially rises slowly with $B_{\mathrm{sw}}$ to a peak of $\sim 6.5 $ keV at $B_{\mathrm{sw}}=10^3 \; \mathrm{nT}$ before falling away sharply to a value of $4.84 \times 10^{-2}$ keV at $B_{\mathrm{sw}} = 10^6 \; \mathrm{nT}$. The analytic expression of equation (\ref{CPCP-nichols}) for CPCP is also plotted for comparison with the results from the SWMF simulations. A discrepancy between the simulation results and the analytic model is apparent: although the general profiles are similar, the SWMF CPCP values are approximately an order of magnitude smaller than the corresponding analytic values. A point which may be explained by the fact that the analytic model does not account for viscous interactions at the magnetopause boundary \citep{nichols2016}. 

Fig. \ref{fig:stacked_plots_exo_lhs_rhs}(c) shows the maximum FAC density as a function of $B_{\mathrm{sw}}$. The magnitude of the FAC density in the analytic model is proportional to the CPCP and Pedersen conductance, given by

\begin{equation}
j_{\|i} \propto \frac{\Sigma_{\mathrm{P}} \Phi_{\mathrm{CPC}}}{R_{\mathrm{P}}},
\end{equation}
and the full details of this relation can be found in \citet{nichols2016}. There is reasonable agreement between the SWMF and analytic results, with the saturation and turnover occurring at $B_{\mathrm{sw}} = 10^3 \; \mathrm{nT}$ for both, although the simulation results peak at a slighter higher value ($\sim 197 \; \upmu \mathrm{A \,m^{-2}}$)  than the analytic results ($\sim 40 \; \upmu \mathrm{A \,m^{-2}}$).  Integration of FAC density over a hemisphere (using equation \ref{itot}) demonstrates that the total ionospheric current also slowly increases wih $B_{\mathrm{sw}}$ up to a peak of $2.1 \times 10^{10} \; \mathrm{A}$ at $B_{\mathrm{sw}} = 10^3 \; \mathrm{nT}$, with a general profile similar to that of CPCP (Fig. \ref{fig:stacked_plots_exo_lhs_rhs}(a)).  However, the absolute values of $I_{\mathrm{tot}}$ exceed those from the analytic model of \citet{nichols2016}. Auroral radio power, calculated using equations (\ref{pow-swmf}) and (\ref{rad-pow}), is plotted in in Fig. \ref{fig:stacked_plots_exo_lhs_rhs}(d), and shows a peak value of $4.5 \times 10^{14}\; \mathrm{W}$ at $B_{\mathrm{sw}} = 10^{3}\; \mathrm{nT}$, a value approximately four orders of magnitude greater than equivalent peak ECMI emission from Jupiter's aurora.

Figs. \ref{fig:stacked_plots_exo_lhs_rhs}(e) -- (h) show the same parameters plotted as a function of Pedersen conductance for a fixed IMF strength of $B_{\mathrm{sw}} = 10^4 \; \mathrm{nT}$.  The results show that the total maximum FAC density, total current, and radio power are all virtually independent of Pedersen conductance for the SWMF simulations, since the low Alfv\'{e}n conductance implied by the high IMF strength means that the condition for saturation described in Section \ref{iono-elec}, namely when $\Sigma_{\mathrm{P}} \gg  \Sigma_{\mathrm{A}}$, is now satisfied for low $\Sigma_{\mathrm{P}}$ values.

%%%%%%%%%%%%%%%%%%
\subsection{Variable orbital distance}
%%%%%%%%%%%%%%%%%%

\begin{figure}
    \centering
    \includegraphics[width=1.0\linewidth]{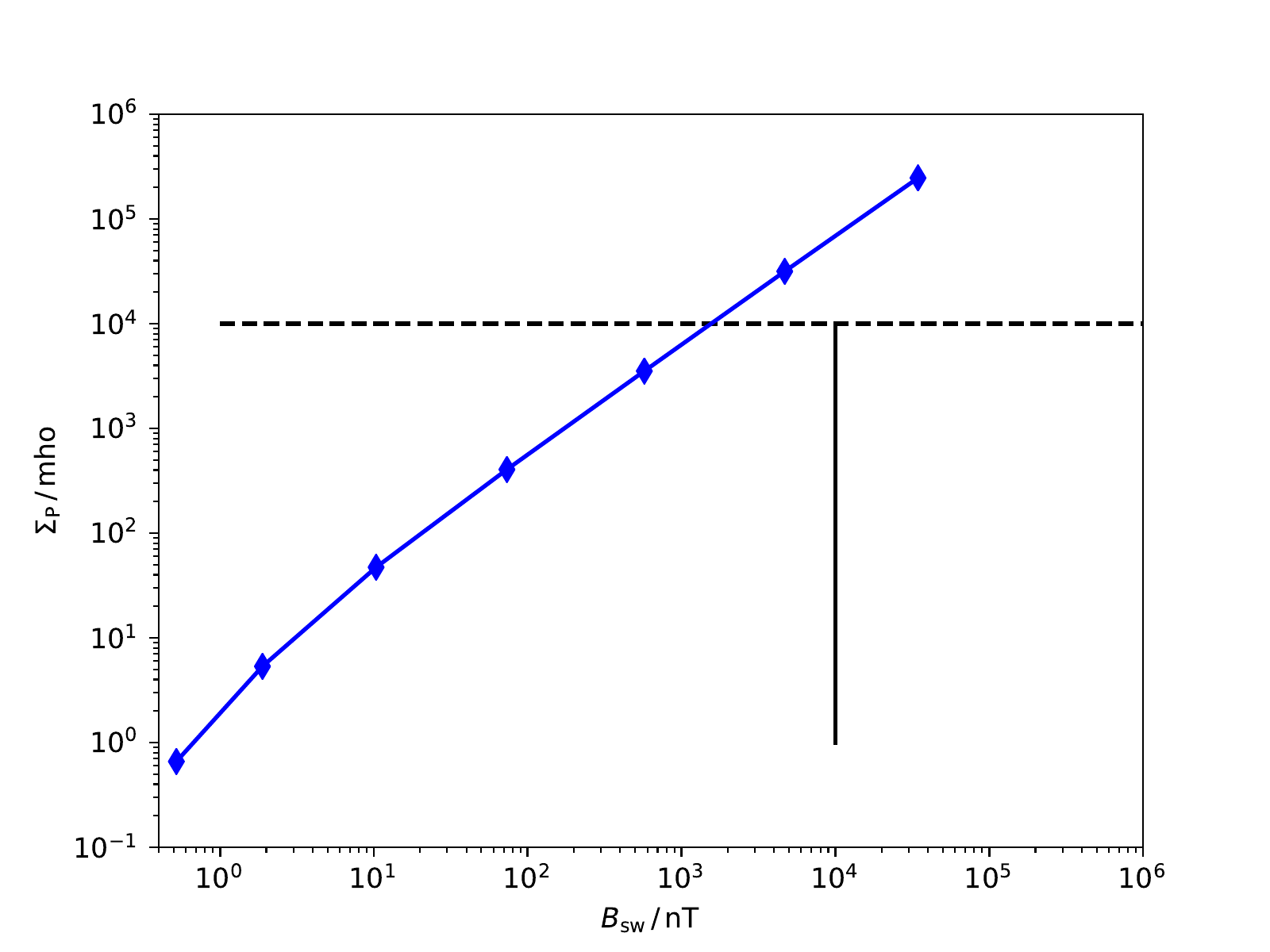}
    \caption{Values of the $\Sigma_{\mathrm{P}}$--$B_{\mathrm{sw}}$ parameter space used in the simulation runs.  The dashed line indicates simulation sets at constant $\Sigma_{\mathrm{P}} = 10^4 \; \mathrm{mho}$, i.e. those runs shown in Fig. \ref{fig:stacked_plots_exo_lhs_rhs}(a)-(d), and the solid black line indicates simulations at constant $B_{\mathrm{sw}} = 10^4 \; \mathrm{nT}$ (Fig. \ref{fig:stacked_plots_exo_lhs_rhs}(e)-(h)). Values representative of varying orbital distances are marked by the blue diamonds.}
    \label{fig:param_cuts}
\end{figure}

Planets at different orbital distances from the host star are not only subject to varying IMF strengths, but also varying stellar wind velocity, density, and Pedersen conductance.  In this section the effects on the CPCP, FACs, and radio power are investigated as functions of orbital distance.  For a Sun-like star, \cite{nichols2016} calculated analytically how stellar wind parameters vary with orbital distance.  Using that work as a reference, spot values of stellar wind velocity, density, IMF strength, and Pedersen conductance were taken at five orbital distances from 2 -- 126 $R_{*}$, and used as inputs for five SWMF simulations. The input parameter values used for each run in the simulation set are summarised in Table \ref{tab:orb_dist}, and Fig. \ref{fig:param_cuts} shows the Pedersen conductance and IMF values used, in relation to the cuts in the parameter space representative of the simulation sets in Section \ref{response}.

\begin{table}
	\centering
	\caption{Input parameter values for orbital distance model runs at a range of orbital distances.}
		\label{tab:orb_dist}
	\begin{tabular}{c |c |c | c | c} 
		\hline	
		$d (R_*)$ & $v_{\mathrm{sw}}\; ( \mathrm{km}\;\mathrm{s}^{-1})$ & $B_{\mathrm{sw}} \; (\mathrm{nT})$ & $n_{\mathrm{sw}} \; (\mathrm{cm}^{-3})$ & $\Sigma_{\mathrm{P}} \; (\mathrm{mho})$ \\
		\hline
		2     & 304.8 & 3.31 $\times 10^{4}$ & 1.38 $\times 10^6$ & 2.46 $\times 10^5$ \\
		5.63  & 225.2 & 3.62 $\times 10^{3}$ & 2.77 $\times 10^4$ & 3.16 $\times 10^4$ \\
		15.87 & 279.8 & 158                  & 1.67 $\times 10^3$ & 3.53 $\times 10^3$ \\
		44.72 & 370.1 & 4.86					& 148				 & 405				 \\
	    126.0 & 453.7 & 10.4					& 15					 & 47					\\
	    	\hline	
	
	\end{tabular}
 
\end{table}

\begin{figure}
    \centering
    \includegraphics[width=1.0\linewidth]{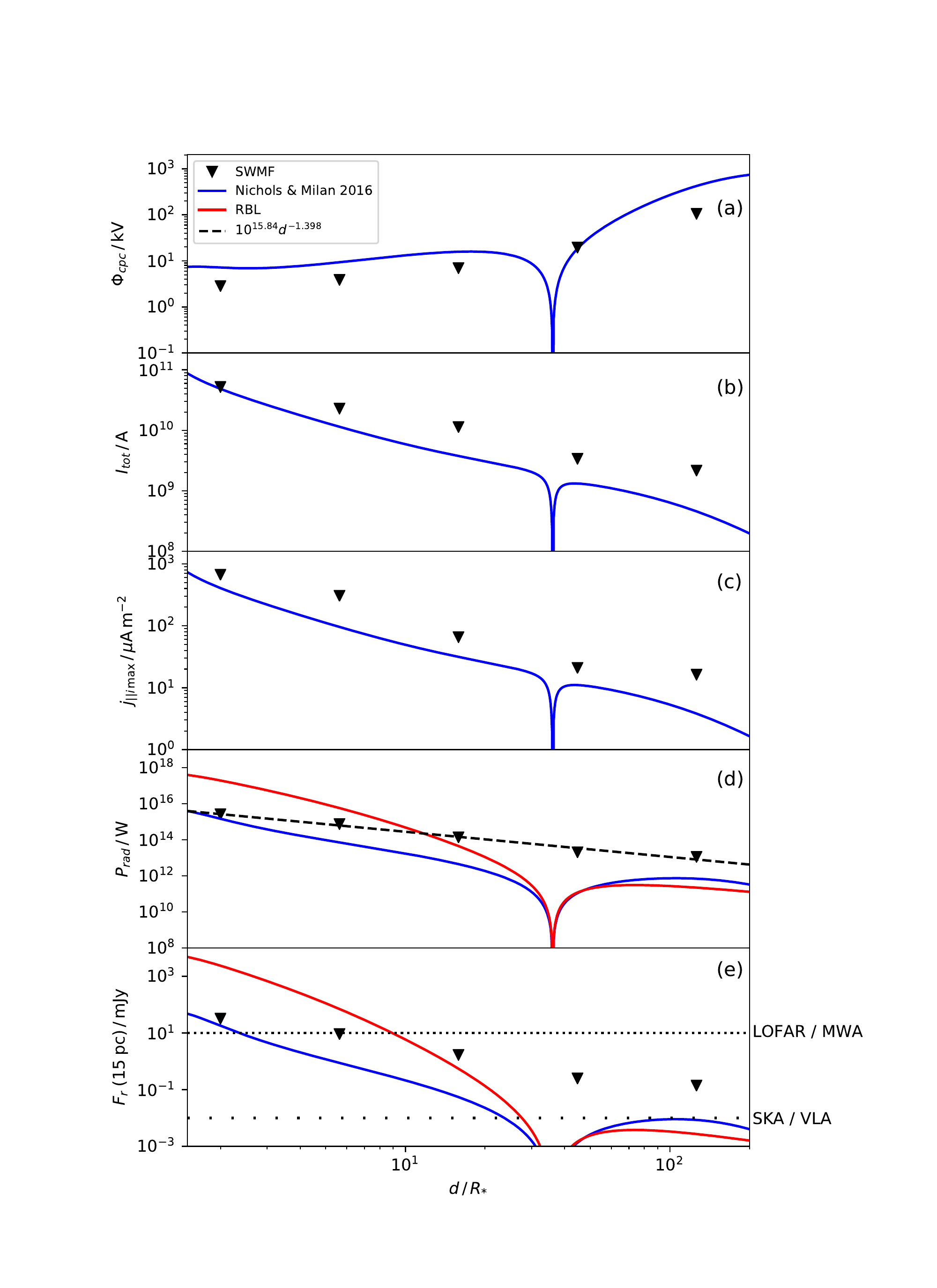}
    \caption[Results as a function of orbital distance]{(a) Cross-polar cap potential, (b) total field-aligned current, (c) maximum ionospheric field-aligned current density, (d) auroral radio power, and (e) spectral flux density as functions of orbital distance $d$ from the central star. Black triangles denote the simulation results, solid blue and red curves indicate comparative analytic functions, and the black dashed line in (d) represents a least squares fit to the SWMF results. Sensitivity thresholds of a number of radio telescopes are marked by horizontal lines in (e).}
    \label{fig:orb_dist}
\end{figure}

Fig. \ref{fig:orb_dist}(a) shows that the CPCP rises from $\sim$ 3 kV in the nominal hot Jupiter orbital distance region, to $\sim$ 106 kV at an orbital distance of $126 \; R_{*}$. Total FAC (Fig. \ref{fig:orb_dist}(b)) falls with increasing orbital distance from a value of $\sim 5 \times 10^{10} \; \mathrm{A}$ at $d = 2 \; R_*$ to $\sim 2 \times 10^9 \; \mathrm{A}$ at $ d = 126 \; R_*$, and a similar trend is observed in the results for maximum FAC density (Fig. \ref{fig:orb_dist}(c)). Figs. \ref{fig:orb_dist}(a) -- (c) show a good agreement between the SWMF results and the analytic results of \citet{nichols2016}. In Fig. \ref{fig:orb_dist}(d) auroral radio power is shown along with analytic results of \citet{nichols2016}, and the Radiometric Bode's law.  The dashed line shows a least-squares polynomial fit to the SWMF results, which yields the  relation $P_{\mathrm{r}} \propto d^{-1.398}$.  \citet{nichols2016}, using a Parker spiral IMF, found that the power varies as $P_{\mathrm{r}} \propto d^{-5/2}$ in the inner orbtial distance range, i.e. before the notch in Fig. \ref{fig:orb_dist}(d), and as $P_r \propto d^{-5/4}$ in the outer orbital distance range, where the resultant IMF is dominated by the perpendicular field component. The relation found in this study of $P_r \propto d^{-1.398}$  therefore lies between the two power laws determined by \citet{nichols2016}. The radio powers in the hot Jupiter orbital range (3 -- 10 $R_*$) of $\sim 10^{15} \; \mathrm{W}$ are commensurate with the peak radio powers seen in the results of Fig. \ref{fig:stacked_plots_exo_lhs_rhs}. Finally, Fig. \ref{fig:orb_dist}(e) plots the spectral flux density $F_{\mathrm{r}}$ calculated using equation (\ref{flux}) at a distance of $s =$ 15 pc, a value chosen since it is apparent that emission from planets significantly beyond this distance would be below the detection threshold of currently available radio telescopes. Horizontal lines in Fig. \ref{fig:orb_dist}(e) indicate the sensitivities of the Murchison Widefield Array (MWA), the Low-Frequency Array (LOFAR), the Very Large Array (VLA), and the Square Kilometre Array (SKA).  The results show that radio flux from hot Jupiters located within 15 pc of the Solar system should be detectable with both the VLA, and in the future with the SKA.

%%%%%%%%%%%%%%%%%%
\section{Discussion and conclusions} 
%%%%%%%%%%%%%%%%%%

The model used in the study contains some inherent limitations due to simplifications made to the realistic dynamics of star-hot Jupiter interactions. The upper atmospheres of hot Jupiters undergo intensive escape due to ionization and radiation heating driven by stellar X-ray and EUV radiation, giving rise to the so-called planetary wind \citep[e.g.][]{yelle2004, munoz2007, murray2009,  koskinen2013}.  This planetary wind is additionally shaped by the gravitational interaction between the star and planet. As can be seen from equations (\ref{massconv}) - (\ref{energyeq2}), gravitational effects and ionisation due to heating and radiation are not included in the MHD model used in this study. Atmospheric escape in the form of a planetary wind would add an additional pressure outward from the planet which would expand the magnetosphere, and we expect this to increase the strength of the radio emission due to an increase in the size of the stellar wind reconnection channel. This study is a  first attempt to establish the M-I coupling dynamics using a 3D MHD model in the framework of existing analytic studies, and as such is primarily intended to investigate the broadbrush effect of IMF strength and Pedersen conductance on the magnetospheric FACs and auroral radio emission. Attempting to incorporate the phenomena of gravitational and ionisation effects immediately may introduce additional unconstrained free parameters which could obscure the intention of this study, namely to isolate the response of the model outputs to IMF strength and ionospheric Pedersen conductance.  A more complete future study should develop this model to incorporate the gravitational and ionization effects on the star-planet interaction and resulting FACs and radio emission.

One of the most notable features of the results shown in Fig. \ref{fig:stacked_plots_exo_lhs_rhs} is the approximately order of magnitude discrepancy between the numerical SWMF results and analytic model for the CPCP.  A possible explanation for this is suggested by \cite{ridley2004}, who remark that the magnetospheres studied in MHD simulations are, by nature, MHD magnetospheres.  Despite accurately depicting the general shape of the pressure distribution, the magnitude is typically underestimated by approximately an order of magnitude at the inner magnetosphere.  This underestimation is a result of the lack of energy discretization in the MHD simulation, meaning that the code is unable to model high energy particles.  As the FAC densities in the simulations presented here are substantially higher than those typically encountered in similar SWMF modelling of Earth magnetosphere, the associated electrons also have correspondingly higher energies.  Hence, an inability to model high energy electron may have more impact to this study than to studies involving less energetic particles at Earth, and thus provide a plausible explanation for the order of magnitude discrepancy seen between numerical and analytic results.

The saturation of field-aligned currents and radio power at relatively low ionospheric Pedersen conductances implies that variations in this parameter may be largely inconsequential at hot Jupiters, where anticipated conductance values are of the order $10^4$ -- $10^5 \; \mathrm{mho}$.  Analytic modelling finds that Pedersen conductance affects the point at which CPCP, and therefore radio power, saturates, but this effect is not apparent in the results from SWMF simulations, where radio power peaks at a point which appears to be independent of Pedersen conductance.  %Low conductance values ($< 10 \; \mathrm{mho}$) would appear to prohibit detectable emission, but such small conductances are unlikely at close-orbiting exoplanets.

Fig.  \ref{fig:stacked_plots_exo_lhs_rhs} shows that as Pedersen conductance is increased, CPCP falls while FAC density remains largely constant for the SWMF results. This reinforces previous findings that the magnetosphere acts as neither a voltage nor current generator \citep{fedder1987, ridley2004}. By considering Ohm's law, $J = \sigma E$, if the magnetosphere is acting as a current generator, where $J$ is constant, then increasing $\sigma$ would result in a linear increase of $E$.  However, Fig. \ref{fig:stacked_plots_exo_lhs_rhs} shows a constant current but non-linear decreasing CPCP. One the other hand, if the magnetosphere acts as a voltage generator then increasing $\sigma$ would correspond to a linear increase of $J$ for a constant $E$.  Neither of these trends are seen in the SWMF results, and in the analytic results both current and potential are found to change simultaneously with $\Sigma_{\mathrm{P}}$.  The results shown in Fig. \ref{fig:stacked_plots_exo_lhs_rhs} represent spot values at horizontal and vertical cuts along a $B_{\mathrm{sw}}$ -- $\Sigma_{\mathrm{P}}$ plane, and exhibit differences between the analytic and numerical MHD values. However, the values plotted in Fig. \ref{fig:orb_dist} for a realistic variation with orbital distance actually agree remarkably well.

Various scaling laws exist which approximate the planetary magnetic dynamo performance to provide estimations of the magnetic field strengths at hot Jupiters. For example \citet{griessmeier2005, griessmeier2007} predict magnetic moments of hot Jupiters to be approximately 10\% of the Jovian field strength. However, a recent study by \citet{cauley2019} find evidence for hot Jupiter surface magnetic field strengths approximately 10 -100 times larger than those predicted by scaling laws. Therefore, while magnetic field strengths at hot Jupiters may be a fraction of the Jovian value, it is also possible that they exceed the Jovian value. In light of the present high degree of uncertainty in the field regarding hot Jupiter magnetic field strengths determined either from scaling laws or from observations of star-planet interaction, the jovian value is employed in this study as a reasonable benchmark alongside which the jovian values of other presently unconstrained hot Jupiter parameters (i.e. source plasma population number density and thermal energy) may also be employed for consistency. 

% In light of this uncertainty, and to maintain consistency with the uncertainty in the plasma source population thermal energy and number density, we employed the Jovian value as a representative spot value. The planetary magnetic field strength is a significant factor governing the magnetospheric FACs and ultimately the auroral radio emission, suggesting the need for future studies examining a wider range of planetary magnetic field strengths.}

Note that we have not considered here hydrodynamic outflow owing to the strong irradiation from the host star or stellar-planetary tidal interaction, and as such our work should be considered to be a first step toward a self-consistent picture of the magnetospheric dynamics of a hot Jupiter. The escaping planetary wind at hot Jupiters should deform the field lines from their dipolar topology, and under certain conditions may distort the magnetosphere into a magnetodisc configuration \citep{khodachenko2015}. At Jupiter the modification of the planetary magnetosphere into a disc-like topology affects the FAC system and the location and size of the main auroral oval \citep{nichols2011, nichols2015}. Similar effects may occur at hot Jupiters, although the presence of a magnetodisc may not be typical for each hot Jupiter, and the effect depends on a variety of parameters which are beyond the scope of this present paper and would require a dedicated future study.

The use of a constant Pedersen conductance in this work serves as a first approximation to a realistic ionosphere, but future work should examine more plausible conductance models. Exoplanets in close orbit around the host star would likely be tidally locked, meaning that one side of the planet permanently faces the star, and is subject to intense ionising stellar radiation.  This could result in an asymmetric Pedersen conductance pattern, with an ionosphere which is highly conductive on the sub-stellar wind side of the planet and has a low conductance on the opposite side.  Particle precipitation from the auroral currents may also further ionise the atmosphere, amplifying the conductance. Such a self-consistent ionospheric model may modify the result presented in this paper, but the general findings would not be expected to alter significantly. This work may also be extended to incorporate planetary rotation.  Since the simulations presented here were run in a time-independent mode, planetary rotation was not a factor, but with simple modification simulations could be run in a time-accurate mode to investigate the effects of rotation on the field-aligned currents and radio power.  In time-accurate mode it would also be possible to examine different stellar wind configurations from the entirely southward $B_z$ IMF in this work. Hence a more realistic Parker spiral type IMF, or a north-south switching IMF could both be implemented.  A future study to explore planetary magnetic field strengths greater the jovian value employed here is also warranted. The trade-off to consider in doing so is the increase in run-time for the SWMF.

Ultimately, this work is intended to guide observations regarding the feasibility of detecting auroral radio emission from exoplanets. The maximum predicted radio powers in this study of $\sim 10^{14}$ -- $10^{15} \; \mathrm{W}$ are consistent with the findings of \citet{nichols2016}, but, in the hot Jupiter regime, are lower than predicted by studies employing the RBL \citep[e.g.][]{zarka2001}, potentially explaining the lack of detection to date.  A particular finding in this work is that such intense radio emission may only occur for a planet exposed to a narrow range of stellar magnetic field strengths. However, this assumes other stellar wind parameters are unchanging, and the simulations for a range of orbital distances (Fig. \ref{fig:orb_dist}), taking into account the variation of these other stellar wind parameters, show a relation between radio power and orbital distance similar to that found in the analytic work of \citet{nichols2016}. The flux density calculations in this study suggest that auroral emission directly from hot Jupiters in the local stellar neighbourhood ($\leq 15$ pc) may be presently detectable with telescopes such as the VLA, and in the near future with the SKA.  Separating an exoplanetary radio signal from background noise may prove challenging, although modulation of the signal at planetary orbital periods could allow light curves to be folded at that period to aid identification of radio signals. 

%%%%%%%%%%%%%%%
\section*{Acknowledgements}
%%%%%%%%%%%%%%%

ST was supported by an STFC Quota Studentship. JDN was supported
by an STFCAdvanced Fellowship (ST/I004084/1) and STFC
grant ST/K001000/1.

%%%%%%%%%%%%%%%%%%%%%%%%%%%%%%%%%%%%%%%%%%%%%%%%%%



\bibliographystyle{mnras}
\bibliography{refs_copy} % if your bibtex file is called example.bib

%\begin{thebibliography}{99}

%\end{thebibliography}

%%%%%%%%%%%%%%%%%%%%%%%%%%%%%%%%%%%%%%%%%%%%%%%%%%

%%%%%%%%%%%%%%%%% APPENDICES %%%%%%%%%%%%%%%%%%%%%

\appendix

%%%%%%%%%%%%%%%%%%%%%%%%%%%%%
\section{MHD modelling of Earth's ionosphere} \label{appA}
%%%%%%%%%%%%%%%%%%%%%%%%%%%%%

Studies by \citet{ridley2004} and \citet{ridley2007} investigated the influence of variable IMF strength and Pedersen conductance on ionospheric cross-polar cap potential and field-aligned currents for Earth. Since this paper investigates similar effects, but at much larger scales and magnitudes, in order to test the method used and to validate the results of this work, the key results from \citet{ridley2004} and \citet{ridley2007} were first reproduced and compared with results from the original studies.

\begin{figure}
    \centering
    \includegraphics[width=1\linewidth]{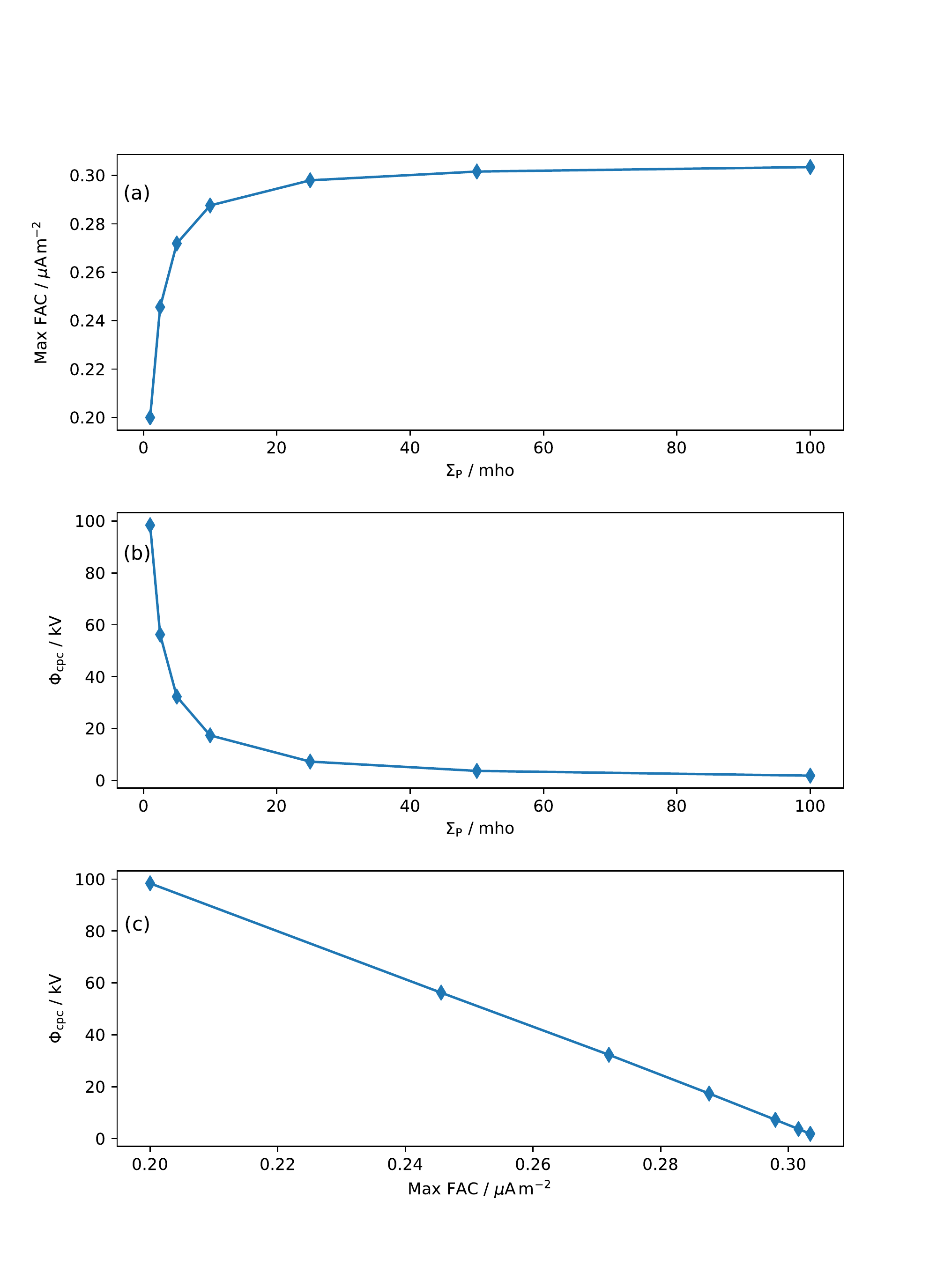}
    \caption[Reproduction of \citet{ridley2004} results]{Results from this work of (a) Maximum ionospheric field-aligned current density and (b) cross-polar cap potential as  functions of Pedersen conductance. (c) Cross-polar cap potential versus the maximum ionospheric field-aligned current density. The diamonds indicate the values at which MHD simulations were run.}
    \label{fig:ridley_repo}
\end{figure}

\begin{figure}
    \centering
    \includegraphics[width=1\linewidth]{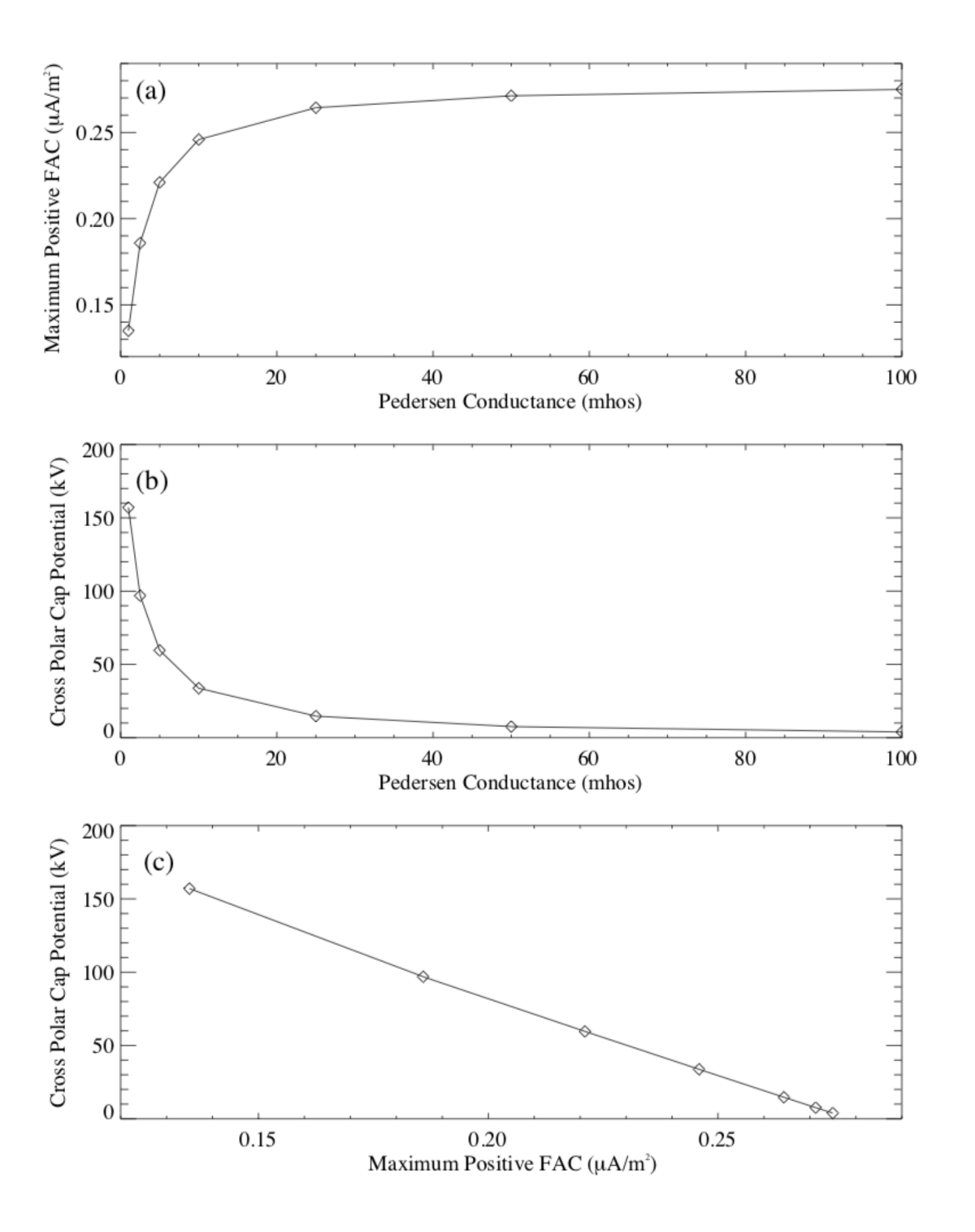}
    \caption[Results from \citet{ridley2004}]{Results from \citet{ridley2004} equivalent to those from \ref{fig:ridley_repo} } 
    \label{fig:ridley}
\end{figure}

\begin{figure}
    \centering
    \includegraphics[width=1\linewidth]{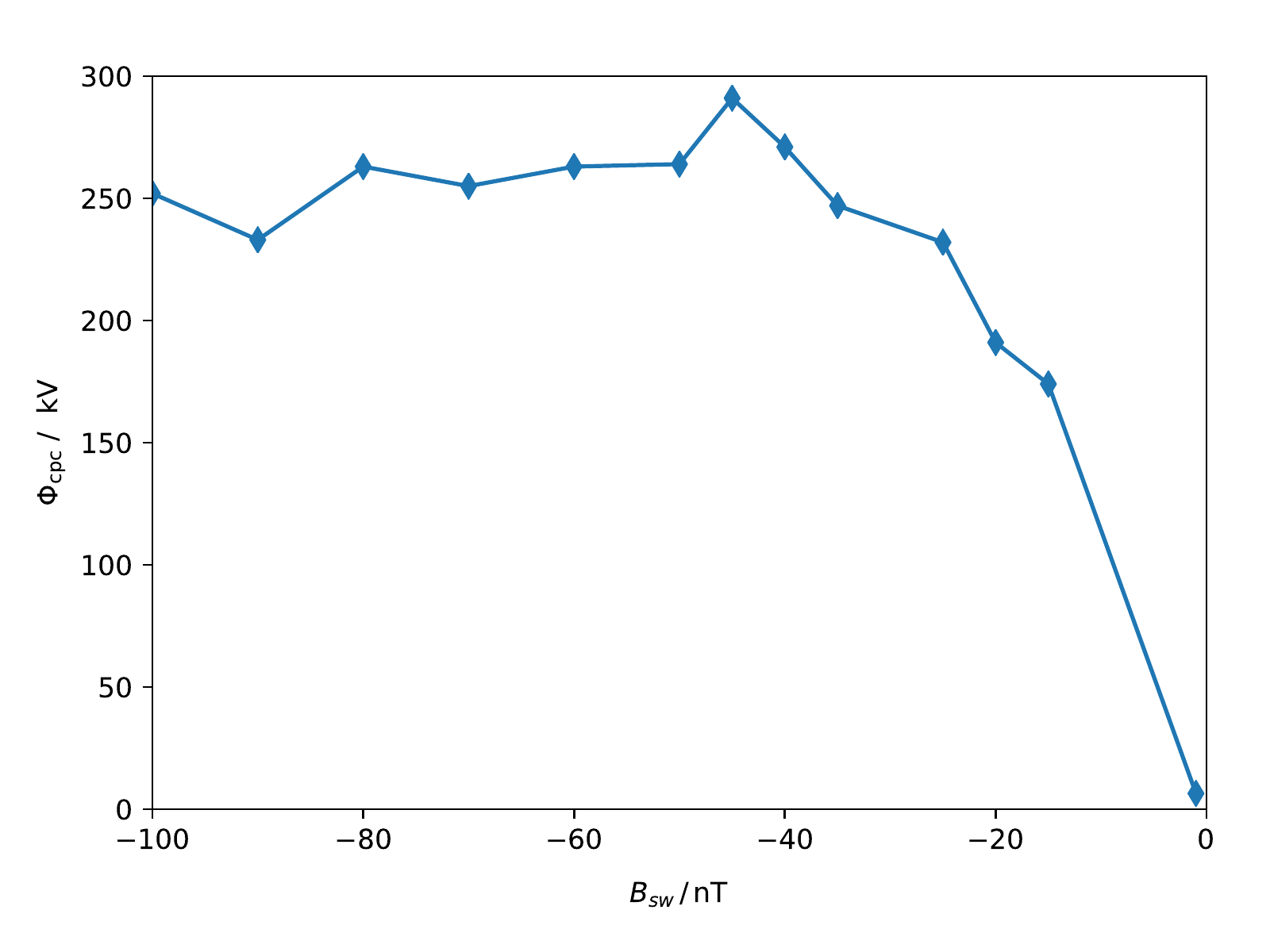}
    \caption[Reproduction of \citet{ridley2007} results]{Ionospheric cross-polar cap potential as a function of IMF strength. Diamonds indicate values at which MHD simulations were run.}
    \label{fig:ridley_repo_sw}
\end{figure}

\begin{figure}
    \centering
    \includegraphics[width=1\linewidth]{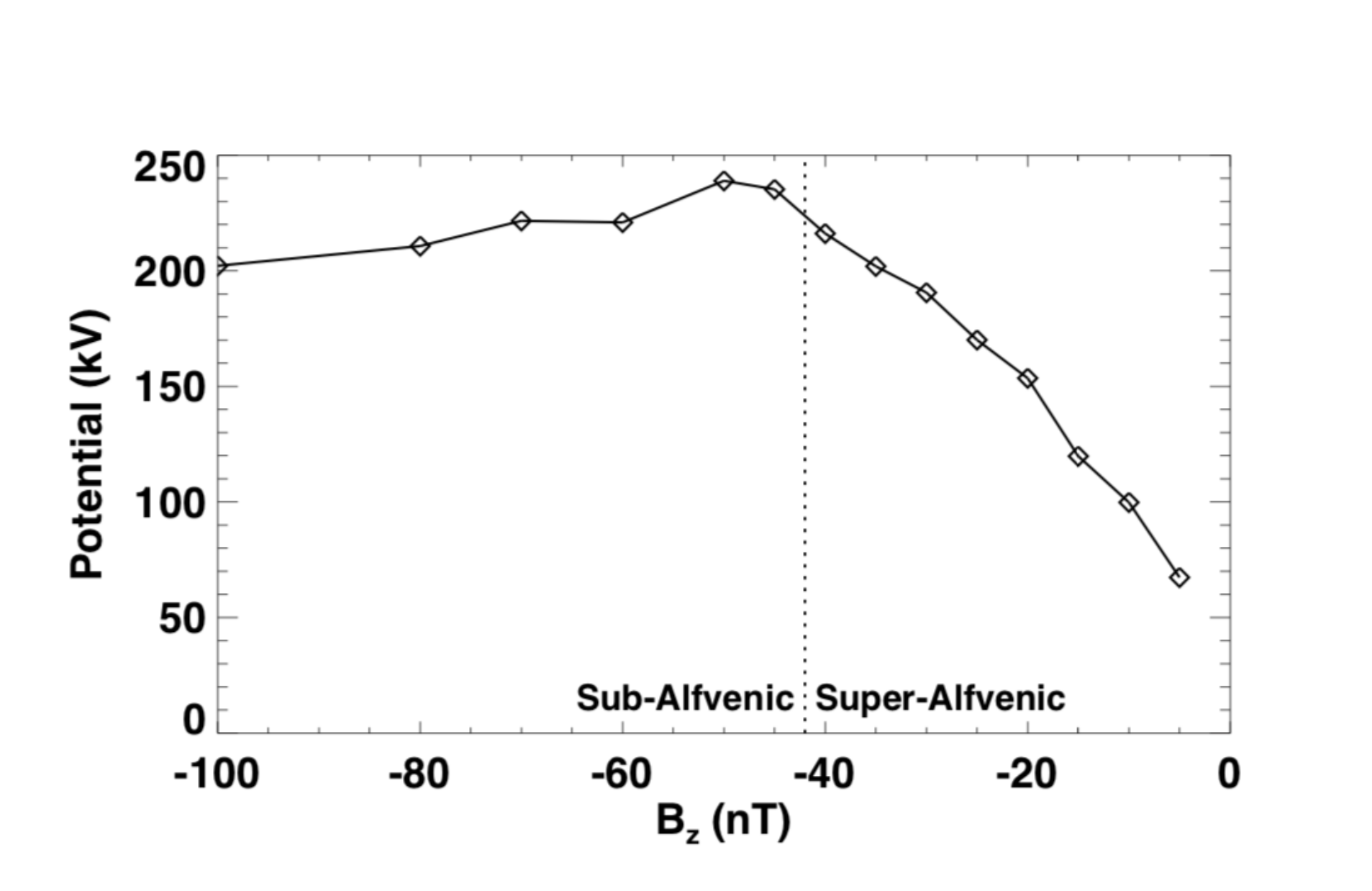}
    \caption[Results from \citet{ridley2007}]{Results from \citet{ridley2007} equivalent to those of \ref{fig:ridley_repo_sw} }
    \label{fig:ridley_sw}
\end{figure}

Simulations using the SWMF were initialised with the terrestrial radius, dipole field strength, and plasma density, as per \citet{ridley2004, ridley2007}. Adaptive mesh refinement of the computational grid was performed in a manner similar to that described by \citet{ridley2007}.  Fig. \ref{fig:ridley_repo} shows the results for the ionospheric field-aligned current density and cross-polar cap potential as functions of Pedersen conductance, which compare closely with the results of \citet{ridley2004} (Fig. \ref{fig:ridley}).  Fig. \ref{fig:ridley_repo_sw} plots cross-polar cap potential as a function of IMF strength, and again the results are in reasonable agreement with those of \citet{ridley2007} (Fig. \ref{fig:ridley_sw}), thus validating the technique employed in this study. 

%
%%%%%%%%%%%%%%%%%%%%%%%%%%%%%%%%%%%%%%%%%%%%%%%%%%

% Don't change these lines
\bsp	% typesetting comment
\label{lastpage}
\end{document}